\documentclass[twocolumn,aps,preprintnumbers,floatfix,amssymb]{revtex4}

\usepackage{graphicx}

\begin{document}

\title{What has NMR taught us about stripes and inhomogeneity?}

\author{Philip M. Singer, Allen W. Hunt, Agneta F. Cederstr\"{o}m}
\affiliation{Department of Physics and Center for Materials Science and 
Engineering, Massachusetts Institute of Technology, Cambridge, MA 02139, USA}

\author{Takashi Imai}
\affiliation{Department of Physics and Astronomy, McMaster University, 
Hamilton, ON L8S-4M1, Canada $^{*}$ E-mail: imai@mcmaster.ca,}
\affiliation{Department of Physics and Center for Materials Science and 
Engineering, Massachusetts Institute of Technology, Cambridge, MA 02139, USA}


\begin{abstract}
The purpose of this brief invited paper is to summarize what we have 
(not) learned from NMR on stripes and inhomogeneity in 
La$_{2-x}$Sr$_{x}$CuO$_{4}$. We explain that the reality is far more 
complicated than generally accepted.    

Keywords: La$_{2-x}$Sr$_{x}$CuO$_{4}$; stripe phase; inhomogeneity
\end{abstract}

\maketitle
(*) Present and permanent address.

Neutron scattering experts generally declare everything {\it static} as soon as fluctuations slow 
down below the frequency threshold of $10^{11}$ Hz for elastic 
scattering measurements. In NMR lineshape measurements, the separation time between 
RF excitation pulses is typically 10 $\mu$s, which sets the 
frequency scale of the {\it shutter speed} of the NMR picture at $10^{5}$Hz.  
In our quest to capture the truly static stripe phase, we 
developed wide-frequency zero-field NMR techniques with a top-loading 
He$^{3}$ cryostat \cite{Hunt2}. Our measurements in 
La$_{1.8-x}$Eu$_{0.2}$Sr$_{x}$CuO$_{4}$ and La$_{2-x}$Ba$_{x}$CuO$_{4}$ 
at $x\sim\frac{1}{8}$ revealed that motional 
narrowing effects average out the hyperfine magnetic fields from Cu 
spins even at 350 mK. Alas, {\it static stripes} are not really 
static even at 350 mK. The observed 
NMR intensity at 350 mK 
accounts for essentailly 100 percent of Cu 
nuclear spins from the sample. This means that the NMR relaxation 
rates $1/T_{1,2}$ 
governed by spin fluctuations at the NMR frequency ($\sim 10^{7}$Hz) 
are small enough to detect the signals, which implies that the 
majority of the 
spectral weight of the spin 
fluctuations has slowed down to $10^{5-6}$Hz levels.  

The dynamic nature of stripes is even 
stronger in La$_{2-x}$Sr$_{x}$CuO$_{4}$ ($x\sim\frac{1}{8}$), 
even though stripes are frequently and erraneously quoted as {\it static} below 
20 K based on elastic neutron scattering data. We 
managed to detect a narrow, Zeeman-perturbed zero-field NMR 
at 350 mK in La$_{1.885}$Sr$_{0.115}$CuO$_{4}$ \cite{Hunt3}. The narrow 
lineshape is a consequence 
of motional narrowing. Moreover, the integrated intensity corresponds 
to only a few percent of the sample.  This implies that the 
majority of Cu nuclear spins are still under the influence of relatively 
fast ($>10^{6}$Hz) fluctuations, and are hence undetectable. 

The dynamic nature of stripes leads 
to an unfortunate consequence; what NMR observes in the {\it static} 
stripe phase is not what one would normally conceive as stripes, but
slowly fluctuating magnetic entities blurred by their motion.  
In order to investigate the slowing of stripes 
through NMR techniques this forces us to rely on
a somewhat indirect method which consist of measuring the Cu NQR/NMR wipeout 
of intensity \cite{Hunt1}. The 
wipeout fraction $F(T)$ is the fraction of the Cu nuclear spins 
that become undetectable due to fluctuating stripes. As detailed in 
\cite{Hunt2}, when the fluctuation frequency falls between 
$10^{11}$Hz and $10^{7}$Hz, very fast NMR 
relaxation rates prevent us from detecting Cu NQR/NMR signals. Effectively, 
$F(T)$ is the volume fraction of the {\it segments} in the CuO$_{2}$ plane 
which fluctuate in the aformentioned frequency range. Quite 
interestingly, $F(T)$ in Nd co-doped samples closely follows the neutron/x-ray scattering 
intensity arising from charge order at 
$x=\frac{1}{8}$\cite{Hunt2,Hunt3,Hunt1,Singer1}.  
This led us to equate $F(T)$ in 
the superconducting regime of La$_{2-x}$Sr$_{x}$CuO$_{4}$ with the volume 
fraction of charge-ordered segments where stripe fluctuations have 
slowed \cite{Hunt1}. The onset temperature $T_{NQR}$ for $F(T)$ 
decreases with increasing $x$\cite{Hunt1,Singer1} (see Fig. 1). Subsequent studies 
\cite{Singer1,Hunt2} showed that 
$T_{NQR}$ precisely agrees with the onset of charge order 
$T_{charge}$ for 
$x>\frac{1}{8}$. This supports our physical picture for $x>\frac{1}{8}$.  
As it turned out, however, $T_{NQR}$ is always higher than 
$T_{charge}$ in the lower Sr doping 
range $x<\frac{1}{8}$ of Nd co-doped samples.
Instead, it is the inflection point of the curvature in the $T$ 
dependence of $F(T)$ that agrees with $T_{charge}$\cite{Hunt2}.  Moreover, subsequent 
x-ray scattering efforts resulted in no hard 
evidence for charge order in La$_{2-x}$Sr$_{x}$CuO$_{4}$ without Nd 
co-doping.  We noted from the very beginning \cite{Hunt1} that the monotonic increase 
of $T_{NQR}$ below $x=\frac{1}{8}$ was {\it counterintuitive}, and we 
were puzzled by these 
newer revelations for some time \cite{Hunt2}.  There must be 
something else involved in the mechanism of $F(T)$ below $T_{NQR}$ 
down to $T_{charge}$ for $x<\frac{1}{8}$. We will come 
back to this point below.

The fact that some Cu 
nuclear spins are observable while others are wiped-out implies a
highly inhomogeneous nature of the CuO$_{2}$  
planes, i.e. that the fluctuation frequency is different position by 
position. This {\it glassy} nature of the stripes led to 
significant confusion in the NMR community. A major source of confusion 
is that from old days everybody in the NMR community knew that NMR data in
La$_{2-x}$Sr$_{x}$CuO$_{4}$
showed a variety of signatures for an electronic inhomogeneity, and distinguishing 
this intrinsic electronic inhomogeneity 
from the inhomogeneous 
magnetism arising from the glassy slowing of stripes is 
not a straightforward task. Some authors even claim that all NMR 
anomalies including wipeout effects may be 
understood based on an analogy with conventional spin 
glass without invoking any spatially coherent nature for the 
stripes \cite{Los}.
Prior to \cite{Los}, we had already 
pointed out in \cite{Hunt1} the importance of the similarity with NMR 
wipeout effects in simple Cu 
metal with dilute magnetic Fe impurity spins. The whole point of Hunt 
{\it et al.} \cite{Hunt1} is that the missing Cu NQR/NMR signal intensity at 
low temperatures in the underdoped regime of La$_{2-x}$Sr$_{x}$CuO$_{4}$, 
which had been attributed to the spatially incoherent slowing of the spin dynamics 
caused by localized holes since the early 90's prior to the 
discovery of stripes by one of us \cite{Imai}, turned out to have some 
hidden information about slowing stripes, and indeed they 
have \cite{Hunt2}. The Los Alamos paper \cite{Los} missed this point. 

In any case, these confusions concerning the inhomogeneity led us to critically 
reexamine the issue of electronic inhomogeneity, more specifially, the 
validity of the assumption that CuO$_{2}$ planes are electronically 
homogeneous even in alloyed high $T_{c}$ cuprates. To make a long story short, 
our measurement of $1/T_{1}$ as a function of temperature {\it and frequency within 
each Cu NQR lineshape} indicates that the local hole concentration 
in La$_{2-x}$Sr$_{x}$CuO$_{4}$ deviates significantly from nominal 
$x$, whether the sample is a poly-crystalline or high-quality single crystal \cite{Singer2}.
This finding clearly raises questions regarding 
theoretical debates of a ``universal electronic phase diagram'', 
including La$_{2-x}$Sr$_{x}$CuO$_{4}$, which are based on 
the assumption that hole doping is homogeneous.
Furthermore, our new result has several implications in our understanding of stripes and the NQR 
wipeout effects. First, the intrinsic electronic inhomogeneity may 
be partially responsible for the 
glassy, inhomogeneous nature of the slowing of stripes observed even at 
$x=\frac{1}{8}$. Second, it naturally explains why the onset temperature 
for wipeout jumps up to $T_{NQR}\sim$ 300 K at 
$x\sim 0.05$. Some segments of CuO$_{2}$ planes 
become nearly undoped below 300 K \cite{Singer2}. For these undoped 
patches, the strong 
short-range N\'{e}el-order results in enhanced NMR relaxation rates 
causing the Cu NQR signal to become undetectable. Whether nucleation of these nearly 
undoped patches is associated with the rotation of the stripes from the
diagonal to vertical direction remains to be seen. Third, similar effects would 
provide a natural account for why wipeout sets in at $T_{NQR}$ above 
$T_{charge}$ below $x=\frac{1}{8}$. The inflection point in the $T$ dependence of $F(T)$ 
corresponds to the temperature where spatially coherent fluctuations 
of glassy stripes finally kick in. 

To summarize, one needs to take into account {\it both} the intrinsic electronic 
inhomogeneity \cite{Singer2} {\it and} the glassy, inhomogeneous slowing of stripes.  
Much of the confusion over La$_{2-x}$Sr$_{x}$CuO$_{4}$ stems from the 
failure by many authors to recognize the importance of both.  


\begin{figure}
\caption{The onset temperature $T_{NQR}$ of Cu NQR wipeout effects in La$_{2-x}$Sr$_{x}$Cu$_{1-y}$Zn$_{y}$O$_{4}$ for $y=0$ ($\circ$), and $y=0.04$ ($\bullet$). 4\% Zn doping suppresses superconductivity completely.}
\end{figure}

\begin{thebibliography}{9}
\bibitem{Hunt2}A.W. Hunt {\it et al.}, Phys. Rev. B {\bf 64}, 134525 (2001).
\bibitem{Hunt3}A.W. Hunt {\it et al.}, unpublished thesis work at M.I.T (2001).
\bibitem{Hunt1}A.W. Hunt {\it et al.}, Phys. Rev. Lett. {\bf 82}, 4300 (1999).
\bibitem{Singer1}P.M. Singer {\it et al.}, Phys. Rev. B {\bf 60}, 15345 (1999).
\bibitem{Singer2}P.M. Singer {\it et al.}, Phys. Rev. Lett. {\bf 88}, 47602 (2002).
\bibitem{Los}N. Curro {\it et al.}, Phys. Rev. Lett. {\bf 85}, 642 (2000).
\bibitem{Imai}T. Imai {\it et al.}, J. Phys. Soc. Jpn. {\bf 59}, 3846 (1990).
\end{thebibliography}
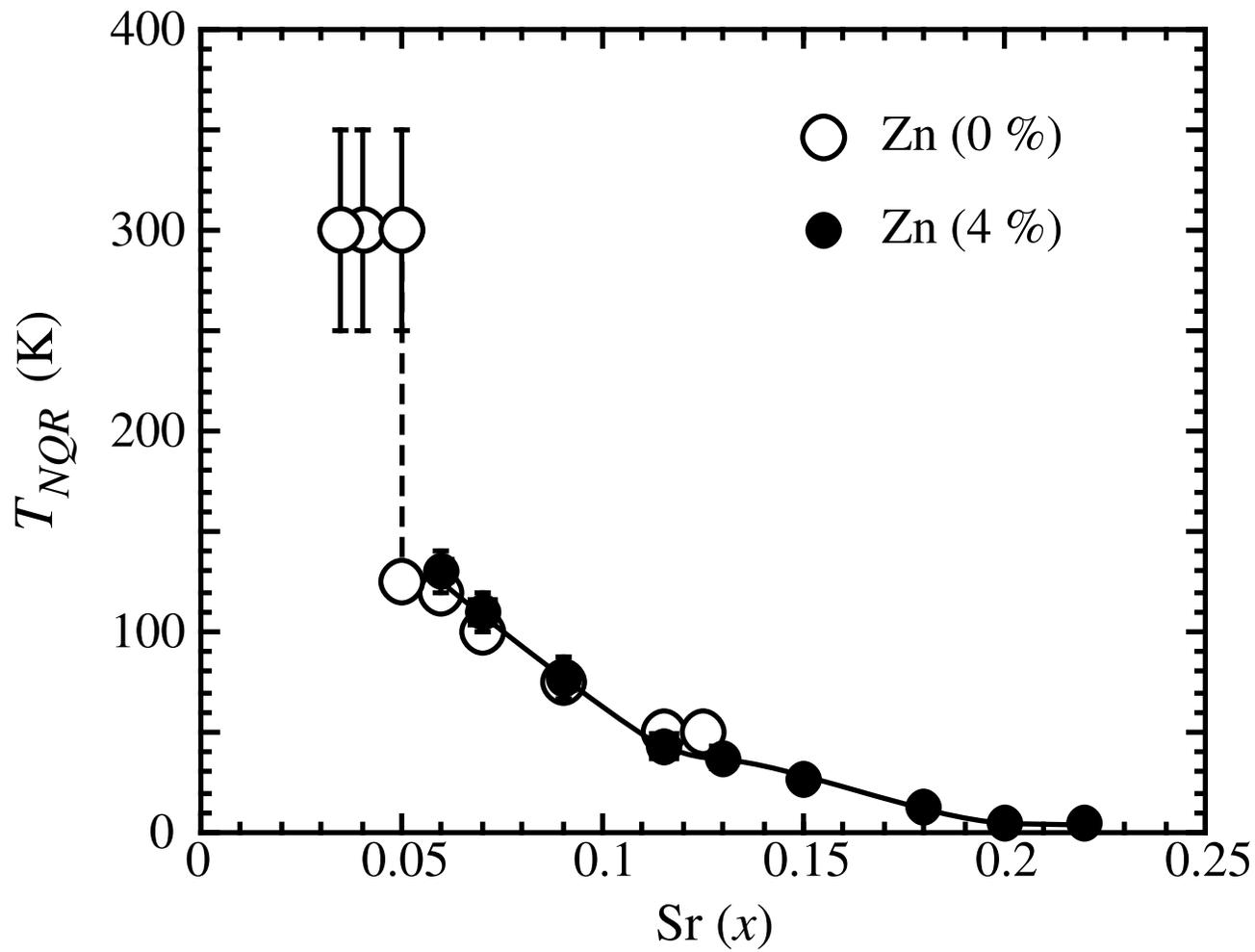
\end{document}